\documentclass[twocolumn,amsmath,amssymb,superscriptaddress,nofootinbib,11pt,a4paper]{revtex4}

\usepackage[utf8]{inputenc}
\usepackage[T1]{fontenc}
\usepackage{CJK} % 如果您需要使用中文
\usepackage{xcolor}
\usepackage{amsfonts}
\usepackage{graphicx}  % Include figure files
\usepackage{dcolumn}   % Align table columns
\usepackage{bm}
\usepackage{float}
\usepackage{braket}
\usepackage{mathrsfs}
\usepackage{subcaption} % 使用 subcaption 而不是 subfigure

\usepackage{mathabx}
\usepackage{siunitx}
\usepackage{comment} % 注释
\usepackage{multirow}
\usepackage{diagbox} % 表格
\usepackage{textcomp}
\usepackage{epstopdf} % 如果需要将 EPS 图形转换为 PDF
\usepackage{wrapfig} % to wrap figure with text
\usepackage{textgreek}
\begin{document}

\title{Energy Extraction via Magnetic Reconnection in Kerr-Sen-AdS$_{4}$ Black Hole: Circular Plasma and Plunging Plasma }

\author{Xiao-Xiong Zeng}
\affiliation{College of Physics and Electronic Engineering, Chongqing Normal University, \\Chongqing 401331, China}

\author{Ke Wang}
\affiliation{School of Material Science and Engineering, Chongqing Jiaotong University, \\Chongqing 400074, China}

\begin{abstract}
{In this paper, we investigate the power and efficiency of energy extraction through magnetic reconnection in the Kerr-Sen-AdS$_{4}$ black hole, with a primary focus on both the circular orbit and  plunging region.
We plot the allowed regions for energy extraction and present corresponding  power and efficiency values, comparing them with those of the Blandford-Znajek (BZ) mechanism.
Our analysis reveals that energy extraction remains feasible even at a spin parameter as low as 0.5, significantly below previously reported thresholds, and the extracted power can exceed that of the BZ process under certain conditions.
Furthermore, the dilatonic scalar charge $b$ and the AdS radius $l$ collectively contribute to lowering the spin threshold for energy extraction.
The energy extraction process in the plunging region was further examined by analyzing the permissible energy extraction region, as well as the corresponding power output and efficiency. One can find that  the energy extraction is possible even at a spin as low as 0.25. Crucially, parameter $b$ actively lowers the energy extraction spin threshold, while parameter $l$  exerts a counteracting effect, impeding such reduction. This trend shows a marked contrast with circular orbit behavior. 
Additionally, both power and efficiency in the plunging region consistently surpass those in the circular orbit region, indicating superior energy extraction capability in plunging region.}
\end{abstract}

\maketitle
%\footnote{corresponding author: %lilifang@imech.ac.cn}

\newpage

\onecolumngrid
\newpage

\section{Introduction}
Magnetic reconnection plays a critical role in astrophysics as it enables the rapid release of magnetic energy, thereby generating substantial power.
It is frequently employed to account for solar flares \cite{1}, coronal mass ejections \cite{2}, and a wide range of other astronomical phenomena \cite{3,4}.
Magnetic reconnection has been extensively studied in magnetohydrodynamics (MHD) \cite{5,6,7}. Recent studies \cite{8,9,10} suggest that spacetime curvature exerts a significant influence on the characteristics of magnetic reconnection, highlighting the need for further investigation into the role of gravitational effects in this process.
The analysis of polarized emission surrounding the supermassive black hole at the center of M87*, as reported by the Event Horizon Telescope collaboration (EHT) \cite{11}, provides further confirmation of the existence of magnetic fields in the vicinity of black holes. This highlights the significance of analyzing magnetic reconnection processes within the theoretical framework of general relativity (GR).

On the other hand,  magnetic reconnection functions as a key mechanism for extracting energy from rotating black holes.
Comisso and Asenjo initially demonstrated that, in the context of a Kerr black hole, the power extracted through this mechanism is generally greater than that achieved via the conventional Blandford-Znajek mechanism \cite{12}. Subsequently, this research was swiftly expanded to encompass other models of rotating black holes \cite{13,14,15,16,17,18,19,20,21,22,23,24,25}. However, this mechanism requires that the spin of the black hole not be excessively low.
For a Kerr black hole with fixed parameters $\xi = \pi / 12$ and $\sigma = 100$, the minimum spin required is 0.85 \cite{12}. In Kerr–de Sitter spacetime, under identical conditions, the minimum spin is found to be no less than 0.7 \cite{17}.
In a rotating regular black hole, under the same conditions and $k=0.1$, the minimum spin is about 0.79 \cite{18}. For a rotating hairy black hole, under the same conditions and $\lambda=0.5,h_0=1.5$, the minimum spin is about 0.86 \cite{19}. 
All results suggest that the black hole spin required for magnetic reconnection varies across different gravitational backgrounds. Therefore, exploring the possibility of energy extraction through magnetic reconnection at lower spin values in alternative gravitational spacetimes constitutes a worthwhile research topic.
In this paper, we examine magnetic reconnection in the vicinity of the Kerr-Sen-AdS$_{4}$ black hole. Our objective is to investigate how the dilatonic scalar charge $ b $ and the AdS radius $ l $ influence the magnetic reconnection process, as well as the associated power and efficiency.
Especially, whether this system demonstrates greater effectiveness in extracting energy through magnetic reconnection compared to previous results. 
We will show that energy extraction is feasible even at a spin value of approximately 0.5 in this system, which represents a more efficient performance compared to previous findings. This improvement can be attributed to the fact that both parameters $b$ and $l$ contribute to lowering the spin threshold required for energy extraction.
 
It is worth noting that in the aforementioned studies, the plasmas all moved along circular orbits outside the photon sphere radius.
Whether magnetic reconnection can be investigated in other regions, such as such as the plunging region, constitutes a highly significant research topic.  In particular, whether the power and efficiency of energy extraction remain consistent across different regions.
In the work of \cite{26}, the scope of energy extraction has been extended to the plunging region. The results indicate that the efficiency of energy extraction in the plunging region is higher than that in the circular orbit region, and the permissible region for energy extraction is also broader.
In the Kerr-Sen-AdS$_{4}$ black hole spacetime, we investigate the magnetic reconnection process occurring within the plunging region, with a focus on analyzing how the dilatonic scalar charge $ b $ and the AdS radius $ l $ influence the efficiency of energy extraction.
It can be observed that even at a spin parameter of approximately 0.25, this black hole is still capable of extracting energy within the plunging region, a threshold lower than that observed in Kerr spacetime. Furthermore, our findings indicate that, in contrast to circular orbits, the parameter $ b $ facilitates a reduction in the spin threshold required for energy extraction within the plunging region, whereas the parameter $ l $ has a suppressing effect.

The remainder of this paper is organized as follows. Section 2 introduces the Kerr-Sen-AdS$_{4}$ spacetime and its characteristics. Section 3.1 describes the magnetic reconnection process in circular orbits. Section 3.2 plots the parameter space for magnetic reconnection in circular orbits. Section 3.3 analyzes the power and efficiency of energy extraction in circular orbits. Section 4.1 introduces the magnetic reconnection process in the plunging region. Section 4.2 analyzes the efficiency and power in the plunging region. We present our conclusions in Section 5.
%%%%%%%%%%%%%%%%%%%%%%%%%%%%%%%%%%%%%
\section{Kerr-Sen-AdS$_{4}$ Spacetime Introduction }
In Boyer-Lindquist(BL) coordinates and geometric units $(c=G=1)$, the Kerr-Sen-AdS$_{4}$ metric is given by \cite{27}:
\begin{equation}
\begin{aligned}
ds^2 &= \frac{-\Delta_r + a^2 \Delta_\theta \sin^2 \theta}{\rho^2} dt^2 + \frac{\rho^2}{\Delta_r} dr^2 + \frac{\rho^2}{\Delta_\theta} d\theta^2  \\
&+ 2 \frac{a \sin^2 \theta}{\rho^2 \Xi} \left[ \Delta_r - \Delta_\theta (r^2 + 2br + a^2) \right] dt \, d\varphi  \\
&+ \frac{\sin^2 \theta}{\rho^2 \Xi^2} \left[ \Delta_\theta (r^2 + 2br + a^2)^2 - a^2 \Delta_r \sin^2 \theta \right] d\varphi^2,
\end{aligned}    
\end{equation}
where
\begin{equation}
\Delta_{r}=\left(1+\frac{r^{2}+2 b r}{l^{2}}\right)\left(r^{2}+2 b r+a^{2}\right)-2 M r,     
\end{equation}
\begin{equation}
\Delta_{\theta}=1-\frac{a^{2}}{l^{2}} \cos ^{2} \theta, \Xi=1-\frac{a^{2}}{l^{2}}, \rho^{2}=r^{2}+2 b r+a^{2} \cos ^{2} \theta, l^{2}=-\frac{3}{\Lambda} ,
\end{equation}
here $a=J / M$ is the angular momentum per unit mass, $b=Q^{2} / 2 M$ is the dilatonic scalar charge, $M$ is the black hole mass, $Q$ is the black hole charge, $\Lambda$ is the negative cosmological constant, $l$ is AdS curvature radius. When $b=0$, the metric reduces to the Kerr-AdS,  when $l=\infty$, it reduces to the Kerr-Sen black hole, and when both conditions are satisfied, it reduces to the Kerr black hole. 

The horizons of the Kerr-Sen-AdS$_{4}$ black hole are determined by $\Delta_{r}=0$. This equation generally has two positive roots: the larger one is the event horizon, and the smaller one is the inner horizon. The boundary of ergosphere is determined by $g_{t t}=0$ , having one root outside the event horizon. We consider the equations of motion for particles on the equatorial plane ($\theta=\pi / 2$) with $M=1$. The radial geodesic equation can be obtained from the Hamilton-Jacobi equation \cite{28}
\begin{equation}
\left(\frac{d r}{d \lambda}\right)^{2}=-\Delta_{r}\left[K+\varepsilon\left(r^{2}+2 b r\right)\right]+\left[\left(r^{2}+2 b r+a^{2}\right) E-a L \Xi\right]^{2}=R(r),\label{4}
\end{equation}
where $\lambda$ is Mino time \cite{29}, related to proper time $\tau$ by $\frac{d \tau}{d \lambda}=\rho^{2}, K$ is the Carter constant \cite{30}, satisfying $K= (a E-L \Xi)^{2}$, $ E$ is the energy, $L$ is the angular momentum. For photons, $\varepsilon=0$,  for massive particles, $\varepsilon=1$. We consider particles on circular orbits in the equatorial plane. The radii of these circular orbits range from infinity down to the photon sphere. The photon sphere radius satisfies
\begin{equation}
R(r)=0, R^{\prime}(r)=0.\label{5}
\end{equation}
Typically, there are two solutions: one for prograde orbits (co-rotating with the black hole) and one for retrograde orbits (counter-rotating). In Fig. 1, we plot the photon sphere radius, boundary of ergosphere, and event horizon $r$ as functions of $a$. Green represents retrograde orbits, purple represents prograde orbits, blue represents the boundary of ergosphere, and red represents the event horizon. We can see that both the cosmological constant and dilatonic scalar charge  reduce the maximum spin and the boundary of the ergosphere compared to the Kerr black hole. The maximum allowed spin for Kerr-Sen-AdS$_{4}$ is less than 1, and the boundary of ergosphere is smaller than 2.

\begin{figure}[!ht]
\centering
\includegraphics[width=0.85\linewidth]{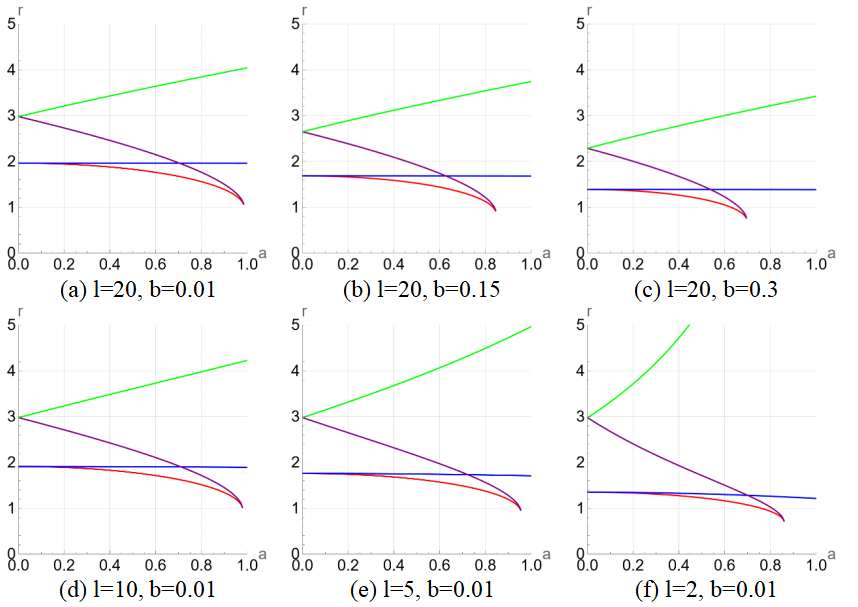}
\caption{Photon sphere radius, boundary of ergosphere, and event horizon $r$ as functions of $a$. The top row shows changes in the four curves for fixed $l$ and varying $b$. The bottom row shows changes for fixed $b$ and varying $l$.}
\label{fig:1}
\end{figure}

From the top row of Fig. 1, we see that for a fixed $l$, as $b$ increases, the allowed maximum spin decreases. This is crucial for enabling energy extraction at lower spins. From the bottom row, we see that for a  fixed $b$, as $l$ decreases, the allowed maximum spin also decreases. Increasing $b$ and decreasing $l$ would increase the deviation from the Kerr black hole. The difference is that in the former case ($b$ increase), the positions of the boundary of ergosphere, event horizon, and photon sphere radius decrease, and the boundary of ergosphere doesn't bend, while in the latter case ($l$ decrease), the positions of the boundary of ergosphere and event horizon decrease, and the boundary of ergosphere bends. Since energy extraction occurs between the event horizon and the boundary of ergosphere, we only need to consider prograde orbits  suggested by Fig. 1. At the photon sphere radius, the particle's energy and angular momentum approach infinity. Therefore, circular orbit energy extraction actually occurs between the photon sphere radius and the boundary of ergosphere. The Keplerian angular velocity for particles on prograde orbits is \cite{28}
\begin{equation}
\Omega=\frac{-\partial_{r} g_{t \varphi}+\sqrt{\left(\partial_{r} g_{t \varphi}\right)^{2}-\left(\partial_{r} g_{t t}\right)\left(\partial_{r} g_{\varphi \varphi}\right)}}{\partial_{r} g_{\varphi \varphi}} .
\end{equation}
This formula comes from the definition of Keplerian angular velocity, $\Omega=\frac{d \varphi / d \tau}{d t / d \tau}$. It can be seen that this formula only involves the angular velocity in the azimuthal direction, not the radial direction, which is characteristic of circular orbits.

\section{Extracting Kerr-Sen-AdS$_{4}$ Black Hole Energy in the Circular Orbit Region }
\subsection{Magnetic Reconnection Process in Circular Orbits}

We use the Zero Angular Momentum Observer (ZAMO) frame for observation \cite{31}. In the ZAMO frame, the metric takes the form  
\begin{equation}
d s^{2}=-d\hat{ t^{2}}+\sum_{i=1}^{3}\left(d \hat{x^{i}}\right)^{2}=\eta_{\mu \nu} d \hat{x^{\mu}} d \hat{x^{\nu}}, 
\end{equation}
where
\begin{equation}
d\hat{t}=\alpha d t, d \hat{x^{i}}=\sqrt{g_{i i}} d x^{i}-\alpha \beta^{i} d t,
\end{equation}
with
\begin{equation}
\alpha=\left(-g_{t t}+\frac{g_{\varphi t}^{2}}{g_{\varphi \varphi}}\right)^{1 / 2}, \beta^{\varphi}=\frac{\sqrt{g_{\varphi \varphi}} \omega^{\varphi}}{\alpha}, \beta^{r}=\beta^{\theta}=0, \omega^{\varphi}=\frac{-g_{\varphi t}}{g_{\varphi \varphi}}.
\end{equation}
Therefore, in the ZAMO frame, the Keplerian velocity is
\begin{equation}
\hat{v}_{K}=\frac{1}{\alpha}\left[\sqrt{g_{\varphi \varphi}} \Omega-\alpha \beta^{\varphi}\right]. 
\end{equation}
We adopt the single-fluid plasma approximation, with the energy-momentum tensor
\begin{equation}
T^{\mu\nu}=p g^{\mu \nu}+w u^{\mu} u^{\nu}+F_{\sigma}^{\mu} F^{\nu \sigma}-(1 / 4) g^{\mu \nu} F^{\alpha \beta} F_{\alpha \beta}, 
\end{equation}
here $w, p, u, F$ are the plasma enthalpy density, pressure, four-velocity, and electromagnetic field tensor, respectively. Assuming the reconnection process is highly efficient, meaning magnetic energy is almost entirely converted to kinetic energy, we can neglect the electromagnetic field tensor part. Using the adiabatic and incompressible nature of the plasma, the energy at infinity per enthalpy for the accelerated and decelerated plasma is approximately \cite{12}
\begin{equation}
\begin{aligned}
e_{ \pm}^{\infty}&=\frac{-\alpha g_{\mu 0} T^{\mu 0}}{w}\\&=\alpha \hat{\gamma}_{K}\left[\left(1+\beta^{\varphi} \hat{v}_{K}\right)\left(1+\sigma\right)^{1 / 2} \pm \cos \xi\left(\beta^{\varphi}+\hat{v}_{K}\right) \sigma^{1 / 2}-\frac{1}{4} \frac{\left(1+\sigma\right)^{1 / 2} \mp \cos{\xi} \hat{v}_{K} \sigma^{1 / 2}}{\hat{\gamma}_{K}^{2}\left(1+\sigma-\cos ^{2}{\xi} \hat{v}_{K}^{2} \sigma\right)}\right],   
\end{aligned}\label{12}
\end{equation}
where $\xi$ is the azimuthal angle of the fluid in the local rest frame, $\sigma$ is the plasma magnetization parameter, and $\hat{\gamma}_{K}$ is the Lorentz factor of $\hat{v}_{K}$ , which can be expressed as 
\begin{equation}
\hat{\gamma}_{K}=(1-\left.\hat{v}_{K}^{2}\right)^{-1 / 2}.
\end{equation}
For a relativistically hot plasma,  $w=4 p$. Similar to the  Penrose process \cite{32}, energy extraction requires satisfying two conditions simultaneously, namely  
\begin{equation}
e_{-}^{\infty}<0, \Delta e_{+}^{\infty}=e_{+}^{\infty}-\left[1-\frac{\Gamma}{4(\Gamma-1)}\right]=e_{+}^{\infty}>0, 
\end{equation}
here $\Gamma$ is the polytropic index, which we take as 4/3. We plot $e_{+}^{\infty}$ and $e_{-}^{\infty}$ in Figs. 2, 3, and 4. Here $r$ is the dominant reconnection point, called the X-point in  \cite{12}.

\begin{figure}[!ht]
\centering
\includegraphics[width=0.6\linewidth]{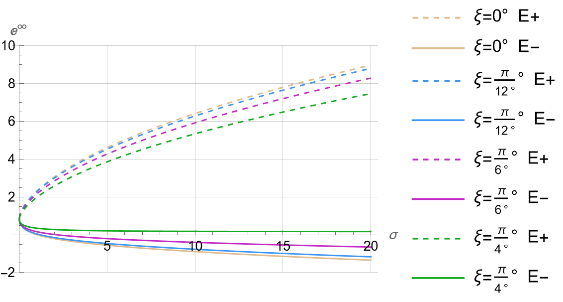}
\caption{The variation of $e_{+}^{\infty}$ and $e_{-}^{\infty}$ with $\sigma$ under different azimuth angles, in which ${a}=0.98, {b}=0.01,l=20,{r}=1.2$.}
\label{fig:2}
\end{figure}

From Fig. 2, it can be seen that $ e_{+}^{\infty}$ is always greater than 0, while $e_{-}^{\infty}$ is almost always less than 0. Therefore, for energy extraction to occur, it mainly depends on whether $e_{-}^{\infty}$ is less than 0. As $\sigma$ increases, $e_{-}^{\infty}$ decreases. As the azimuthal angle increases, $e_{-}^{\infty}$ increases. This behavior is similar to the general case.

\begin{figure}[!ht]
\centering
\includegraphics[width=1.05\linewidth]{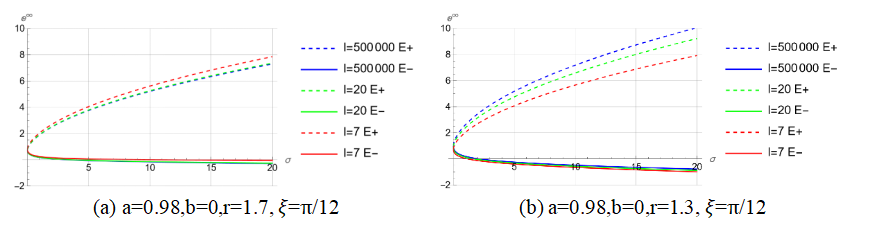}
\caption{The variation of $e_{+}^{\infty}$ and $e_{-}^{\infty}$ with $\sigma$ under different $l$ .}
\label{fig:3}
\end{figure}

From Fig. 3, we see that $e_{+}^{\infty}$ and $e_{-}^{\infty}$ do not vary monotonically with $l$, and their behavior depends on the location of the reconnection point. For smaller $r$, the difference for different $l$ is more obvious.

\begin{figure}[!ht]
\centering
\includegraphics[width=0.6\linewidth]{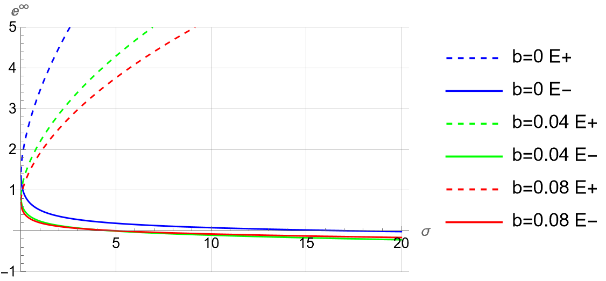}
\caption{The variation of $e_{+}^{\infty}$ and $e_{-}^{\infty}$ with $\sigma$ under different $b$, in which ${a}=0.9, {r}=1.6,l=20,\xi=\pi / 12$. }
\label{fig:4}
\end{figure}

From Fig. 4, we see that as $b$ increases, $e_{-}^{\infty}$ gradually decreases, indicating that energy extraction becomes more favorable. It should be note that $a$ cannot be set to 0.98 here, because $b=0.04$ and $b=0.08$, a naked singularity appears.

\subsection{Parameter Space for  Energy Extraction Via Magnetic Reconnection in Circular Orbits}

We plot the allowed region for energy extraction ($e_{-}^{\infty}<0$) in the $r-a$ plane in Figs. 5-8. The red solid line represents the event horizon, the blue solid line represents the ergosphere, the purple dashed line represents the photon sphere radius, and from left to right, $\sigma=100$, 30, 10, 3.

\begin{figure}[!ht]
\centering
\includegraphics[width=0.99\linewidth]{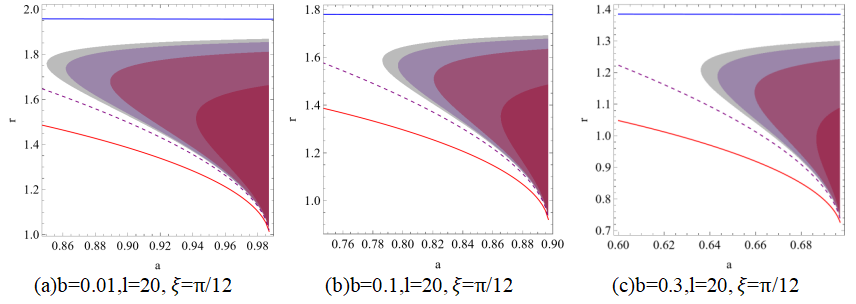}
\caption{ Allowed energy extraction regions for different $b$. }
\label{fig:5}
\end{figure}

From Fig. 5, we see that as $\sigma$ increases, the allowed energy extraction region also increases, which is consistent with   \cite{12}. As $b$ increases, the maximum allowed spin decreases, enabling energy extraction at lower spins. For example, the maximum spin allowed for the black hole in Fig. 5(a) is 0.98753, while in Fig. 5(c) it is 0.6977. The minimum spin allowing energy extraction also decreases, from 0.85 in Fig. 5(a) to 0.64 in Fig. 5(c). Simultaneously, due to the decrease in the ergosphere, event horizon, photon sphere radius, and the position of the reconnection layer also decreases.

\begin{figure}[!ht]
\centering
\includegraphics[width=0.99\linewidth]{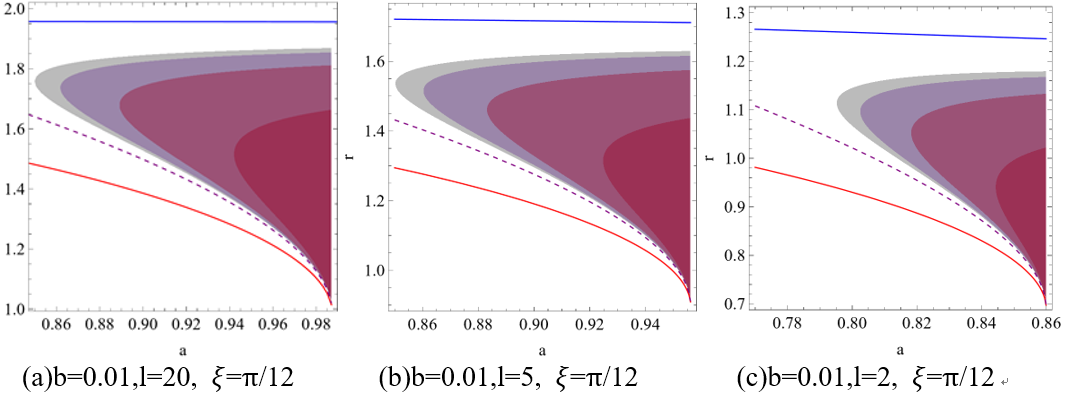}
\caption{Allowed energy extraction regions for different $l$. }
\label{fig:6}
\end{figure}

From Fig. 6, we see that as $l$ decreases, the maximum allowed spin also decreases, enabling energy extraction at lower spins. The minimum spin allows energy extraction   decreases too, analogous to the effect of increasing $b$. Therefore, we conclude that in circular orbits, both parameters $b$ and $l$ promote lowering the spin threshold for energy extraction.

\begin{figure}[!ht]
\centering
\includegraphics[width=0.99\linewidth]{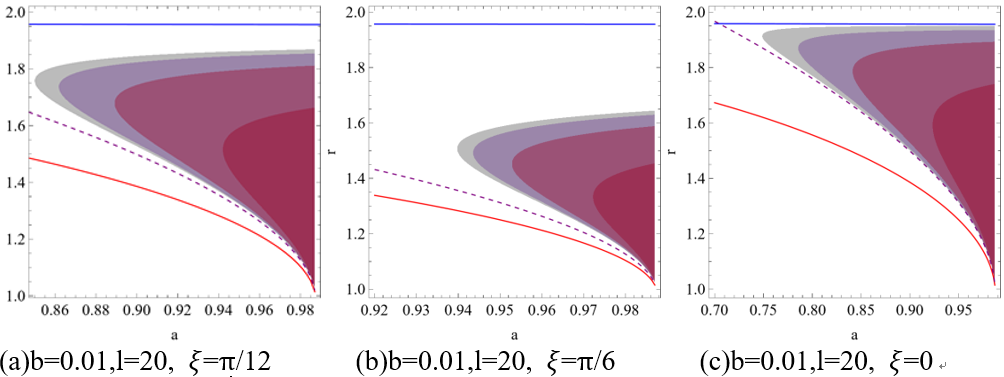}
\caption{Allowed energy extraction regions for different $\xi$. }
\label{fig:7}
\end{figure}

From  Fig. 7, we see that as $\xi$ decreases, the area of the allowed energy extraction region increases, the minimum allowed spin decreases, and the position of the reconnection layer increases. Thereafter, we will not pay attention to the effect of $\xi$ on the magnetic reconnection.  
Without loss of generality,  we choose  $\xi=\pi / 12$.

\begin{figure}[!ht]
\centering
\includegraphics[width=0.4\linewidth]{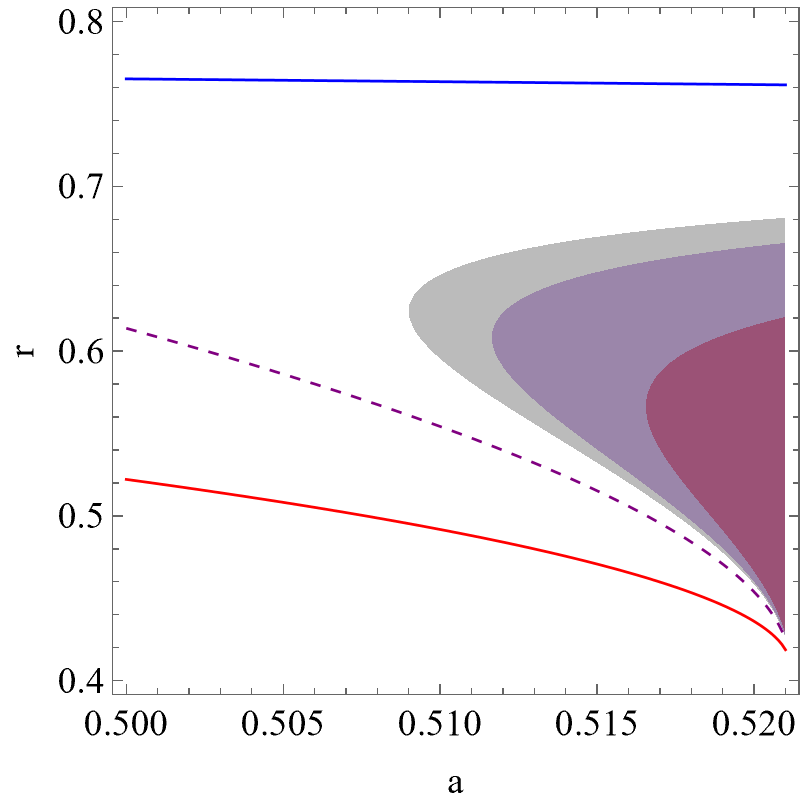}
\caption{Allowed energy extraction region for ${b}=0.36,l=2, \xi=\pi / 12$ }
\label{fig:8}
\end{figure}

In  Fig. 8, we intend to explore how low for the spin $a$ can be as the 
magnetic reconnection occurs. We find for the 
 large $b$ and small $l$, the spin will be lower. The result shows that the allowed spin range for energy extraction is between 0.51 and 0.521, demonstrating that energy extraction is possible at moderate spins.

\subsection{Power and Efficiency of Energy Extraction in Circular Orbits}

Having established the feasibility of energy extraction, we compare the power and efficiency. The energy extraction power is \cite{12}
\begin{equation}
P=-e^{\infty}_- w A_{i n} U_{i n}.\label{15}
\end{equation}
For collisionless conditions,  $U_{i n} \approx 0.1$\cite{33}, for collisional conditions,  $U_{i n} \approx 0.01$\cite{34,35}. We take the first case in this paper. $A_{i n}$ is the cross-sectional area of the inflowing plasma can be expressed  as
\begin{equation}
A_{i n} \sim\left(r_{E}^{2}-r_{p h}^{2}\right), \label{16}
\end{equation}
where $r_{E}$ is the boundary of ergosphere and $r_{p {h}}$ is the photon sphere radius.

\begin{figure}[!ht]
\centering
\includegraphics[width=0.5\linewidth]{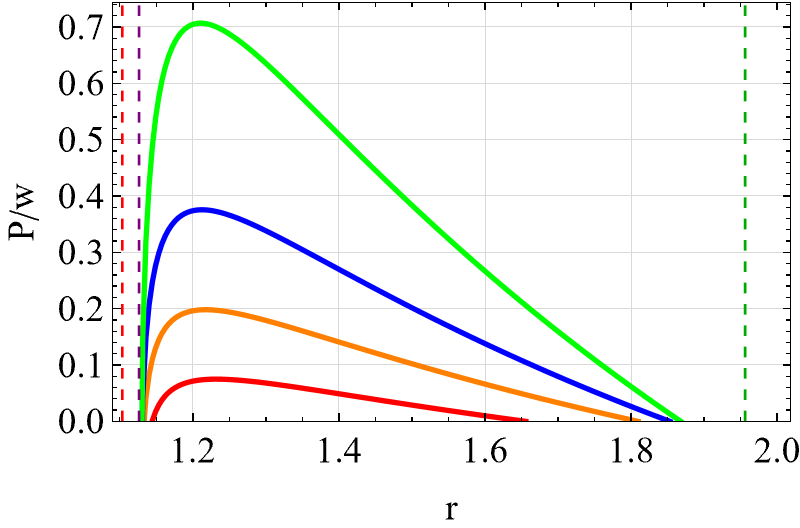}
\caption{Energy extraction power for different $\sigma$, in which  ${b}=0.01,l=20, \xi=\pi / 12, {a}=0.98$. }
\label{fig:9}
\end{figure}

We plot the energy extraction power per enthalpy density $P / w$ with respect to  $r$ in Figs. 9-12. Red dashed, purple dashed, and green dashed lines represent the event horizon, photon sphere, and boundary of ergosphere, respectively. Red solid, orange solid, blue solid, and green solid curves represent $\sigma=3, ~ 10, ~ 30, ~ 100$.

\begin{figure}[!ht]
\centering
\includegraphics[width=0.5\linewidth]{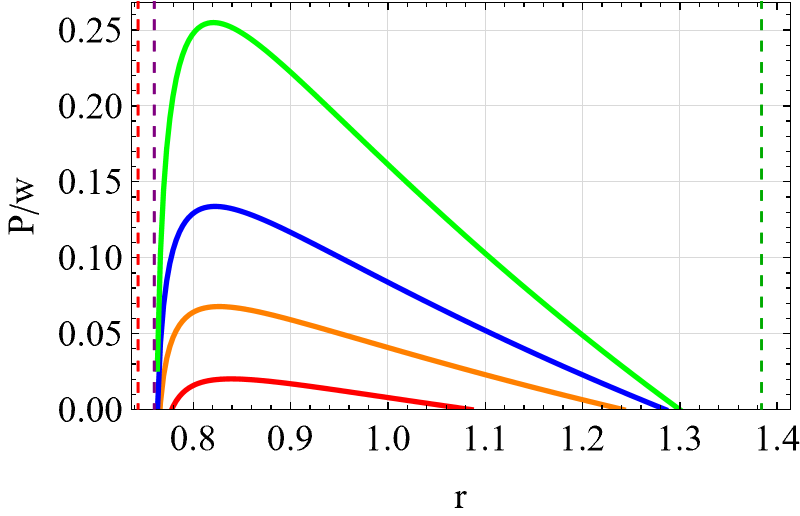}
\caption{Energy extraction power for different $\sigma$, in which   ${b}=0.3,l=20, \xi=\pi / 12, {a}=0.696$. }
\label{fig:10}
\end{figure}

From Fig. 9 we can see that the power starts outside the photon sphere radius, consistent with circular orbit characteristics. The power first increases and then decreases. As $\sigma$ increases, the power increases.
 Comparing Figs. 9 and 10, we see that as $b$ increases, the energy extraction power decreases for the same $\sigma$.  Note that in Figs. 9 and 10, $a$ cannot be the same because the allowed spin of the black hole varies under different $b$ values. This is similar to  \cite{19}, in which  $a$ cannot be the same because the allowed spin of the black hole varies under different $h_0$ values. Comparing Figs. 9 and 11, we see that as $l$ decreases, the energy extraction power also decreases due to the lower spin. Similarly, $a$ cannot be the same because the allowed spin of the black hole varies under different $l$.
Especially, for some special values of $b$ and $l$, which is shown in  Fig. 12, we find the energy extraction power becomes even lower.
\begin{figure}[!ht]
\centering
\includegraphics[width=0.5\linewidth]{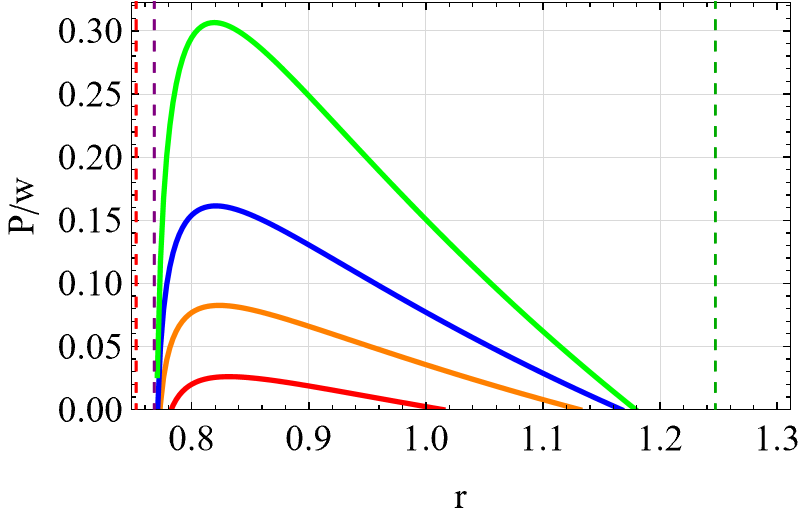}
\caption{Energy extraction power  for different $\sigma$, in which   ${b}=0.01,l=2, \xi=\pi / 12, {a}=0.856$. }
\label{fig:11}
\end{figure}

\begin{figure}[!ht]
\centering
\includegraphics[width=0.5\linewidth]{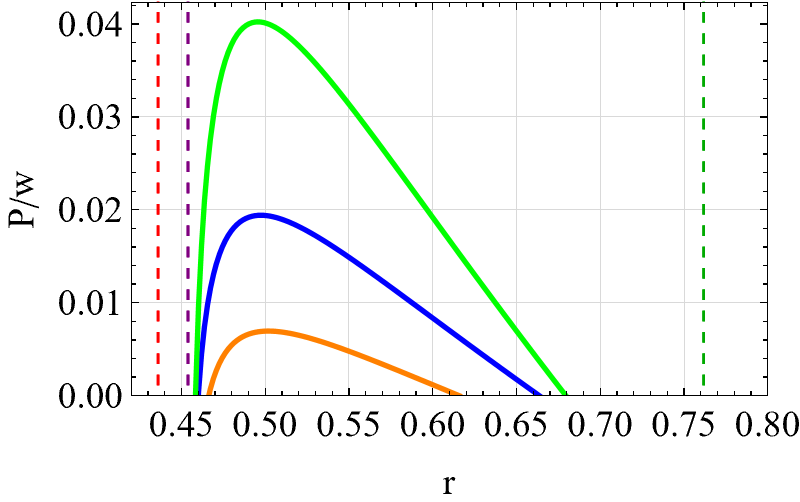}
\caption{Energy extraction power  for different $\sigma$, in which    ${b}=0.36,l=2, \xi=\pi / 12, {a}=0.52$. $b$ and $l$.}
\label{fig:12}
\end{figure}

Next, we intend to discuss the energy extraction efficiency, which is defined as  \cite{12}
\begin{equation}
\eta=\frac{e_{+}^{\infty}}{e_{+}^{\infty}+e_{-}^{\infty}}.\label{17}
\end{equation} 
Since the energy extraction condition requires $e_{-}^{\infty}<0, e_{+}^{\infty}>0$, for energy extraction to occur, the efficiency must be greater than 1. We plot the energy extraction efficiency with respect to  $r$ in Figs. 13-16. We fix $\xi=\pi / 12, \sigma=100$.

\begin{figure}[!ht]
\centering
\includegraphics[width=0.6\linewidth]{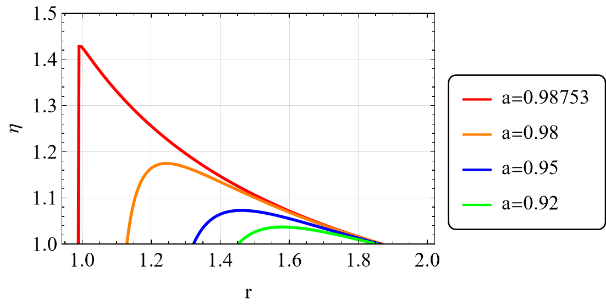}
\caption{ Energy extraction efficiency  for different $a$, in which   ${b}=0.01,l=20$.}
\label{fig:13}
\end{figure}

\begin{figure}[!ht]
\centering
\includegraphics[width=0.6\linewidth]{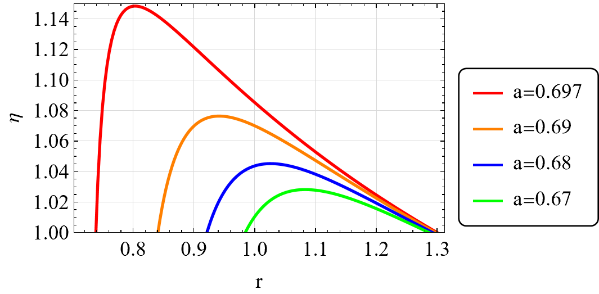}
\caption{Energy extraction efficiency plot for different $a$, in which   \textbf{}  ${b}=0.3,l=20$.}
\label{fig:14}
\end{figure}

In Fig. 13, we plot the  efficiency for different $a$ under the small  $b$. 
Like the power, we find the efficiency first increases and then decreases. As $a$ increases, the efficiency increases. 
For the larger $b$, the  efficiency for different $a$  is shown in in Fig. 14.
Comparing Figs. 13 and 14, we see that as $b$ increases, the energy extraction efficiency decreases due to the lower spin, similar to the power trend.  In Figs. 13 and 14, $a$ cannot be the same because the allowed spin of the black hole varies under different $b$ values. As $b$ is fixed, we also can discuss the effect of $l$ on the energy extraction efficiency.
Comparing Figs. 13 and 15, we see that as $l$ decreases, the energy extraction efficiency also decreases due to the lower spin, similar to the power trend. Especially, for some special values of $b$ and $l$, which is shown in  Fig. 16. Comparing From Fig. 13  and From Fig. 16, we find the efficiency becomes lower, which is also similar to the power trend.

\begin{figure}[!ht]
\centering
\includegraphics[width=0.6\linewidth]{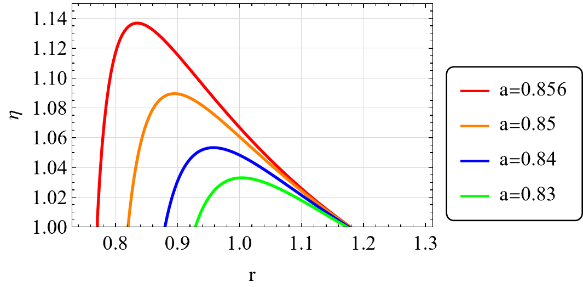}
\caption{Energy extraction efficiency plot for different $a$, in which     ${b}=0.01,l=2$.}
\label{fig:15}
\end{figure}

\begin{figure}[!ht]
\centering
\includegraphics[width=0.6\linewidth]{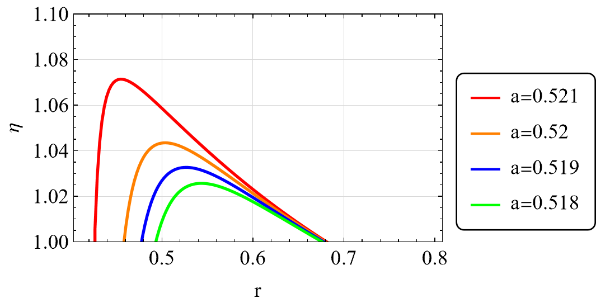}
\caption{Energy extraction efficiency  for different $a$, in which     ${b}=0.36,l=2$.}
\label{fig:16}
\end{figure}

Finally, we compare the energy extraction power ratio. We compare it with the very popular Blandford-Znajek (BZ) mechanism \cite{36}. The BZ energy extraction power is \cite{37}
\begin{equation}
P_{B Z}=\frac{\kappa}{16 \pi} \Phi_{H}^{2}\left(\Omega_{H}^{2}+c_{1} \Omega_{H}^{4}+c_{2} \Omega_{H}^{6}+\mathcal{O}\left(\Omega_{H}^{8}\right)\right), 
\end{equation}
where $\kappa=0.05, c_{1}=1.38, c_{2}=-9.2$ are numerical constants, and $\Phi_{H}$ is the magnetic flux threading one hemisphere of the black hole event horizon, which takes the form 
\begin{equation}
\Phi_{H}=\frac{1}{2} \iint\left|B^{r}\right| \sqrt{g_{\theta \theta} g_{\varphi \varphi}} d \theta d \varphi=\frac{2 \pi\left(r_{+}^{2}+2 b r_{+}+a^{2}\right)}{\Xi} B_{0} \sin (\xi) ,
\end{equation}
where $r_{+}$ is the event horizon radius, $B_{0}=\sqrt{w \sigma}, ~ \Omega_{H}$ is the angular velocity of the event horizon, namely 
\begin{equation}
\Omega_{H}=\left.\frac{-g_{\varphi t}}{g_{\varphi \varphi}}\right|_{r=r_{+}}=\frac{a \Xi}{r_{+}^{2}+2 b r_{+}+a^{2}}.
\end{equation}
Therefore, the power ratio is
\begin{equation}
\frac{P}{P_{B Z}}=\frac{-4 e^{\infty}_- A_{i n} U_{i n} \Xi^{2}}{\kappa \pi \sigma\left(\Omega_{H}^{2}+c_{1} \Omega_{H}^{4}+c_{2} \Omega_{H}^{6}\right) \sin ^{2}(\xi)\left(r_{+}^{2}+2 b r_{+}+a^{2}\right)^{2}}.\label{21}
\end{equation}

From \eqref{21}, it can be seen that as long as the azimuthal angle is sufficiently small, the power ratio can always be made greater than 1, meaning the magnetic reconnection mechanism always has higher power than the BZ mechanism. Without loss of generality, we fix  $\xi=\pi / 12$. Taking parameters for high spin (${b}=0.01,l=20, \sigma=3, {a}=0.98, {r}=1.2$), we obtain a power ratio of 10.3056 > 1, confirming that magnetic reconnection power exceeds BZ power. For example, at low spin and large deviation (${b}=0.36,  l=2,  {a}=0.52, {r}=0.5,  \sigma=7$), the power ratio is 5.56032. Even though the absolute power and efficiency are reduced in this case, the magnetic reconnection power still exceeds the BZ power. This is thanks to the factors $b$ and $l$, which lower the spin threshold and reflect the high efficiency of magnetic reconnection.

\section{ Extracting Kerr-Sen-AdS$_{4}$ Black Hole Energy in the Plunging Region}
\subsection{Magnetic Reconnection Process in the Plunging Region}

Previous studies considered plasma moving on circular orbits outside the photon sphere radius. Here we consider another scenario as in \cite{26}.  That is, the plasma initially moves on a circular orbit outside the Innermost Stable Circular Orbit (ISCO), then plunges inward starting from the ISCO. Circular orbits inside the ISCO are unstable,  any perturbation induces radial velocity, initiating the plunge. The region $r<r_{I}$ (where $r_{I}$ is the ISCO radius, typically larger than the photon sphere radius) is called the plunging region \cite{38}. In the plunging region, because there is radial velocity, the Keplerian velocity formula in \eqref{12}, which only contained the azimuthal component, is no longer applicable. We still use the convenient ZAMO frame. The relationship between the four-velocity in the ZAMO frame and the BL frame is
\begin{equation}
\hat{U}^{\mu}=\hat{\gamma}_{s}\left\{1, \hat{v}_{s}^{(r)}, 0, \hat{v}_{s}^{(\varphi)}\right\}=\left\{\frac{E-\omega^{\varphi} L}{\alpha}, \sqrt{g_{r r}} U^{r}, 0, \frac{L}{\sqrt{g_{\varphi \varphi}}}\right\} ,\label{22}
\end{equation}
where 
\begin{equation}
\left(U^{r}\right)^{2}=\left(\frac{d r}{d \tau}\right)^{2}=\frac{R(r)}{\rho^{4}}, \label{23}   
\end{equation}
\begin{equation}
\rho^{2}\left(\frac{d t}{d \tau}\right)=\frac{E\left(r^{2}+2 b r+a^{2}\right)^{2}-a\left(r^{2}+2 b r+a^{2}\right) L \Xi}{\Delta_{r}}-\frac{\sin ^{2} \theta}{\Delta_{\theta}}\left(a^{2} E-\frac{a L \Xi}{\sin ^{2} \theta}\right),     
\end{equation}
\begin{equation}
E=-\left(g_{t t}+\Omega g_{t \varphi}\right)\left(\frac{d t}{d \tau}\right),    
\end{equation}
\begin{equation}
L=\left(g_{t \varphi}+\Omega g_{\phi \phi}\right)\left(\frac{d t}{d \tau}\right),    
\end{equation}
The conserved quantities in the plunging region are the energy $E$ and angular momentum $L$, at the ISCO, we label them as 
\begin{equation}
E_{I}=E\left(r_{I}\right), L_{I}=L\left(r_{I}\right). \label{27}
\end{equation}
Substituting \eqref{27} into \eqref{23}, and using $R(r)$ satisfying \eqref{4} with $\varepsilon=1$, we get
\begin{equation}
U^{r}=-\frac{1}{\rho^{2}} \sqrt{-\Delta_{r}\left[\left(a E_{I}-L_{I} \Xi\right)^{2}+\left(r^{2}+2 b r\right)\right]+\left[\left(r^{2}+2 b r+a^{2}\right) E_{I}-a L_{I} \Xi\right]^{2}}.\label{28}
\end{equation}
The negative sign indicates inward motion. Substituting \eqref{28} into \eqref{22} gives $\hat{v}_{s}^{(r)}, \hat{v}_{s}^{(\varphi)}$, and\\ $\hat{v}_{s}=\sqrt{\left(\hat{v}_{s}^{(r)}\right)^{2}+\left(\hat{v}_{s}^{(\varphi)}\right)^{2}}$,$\hat{\gamma}_{s}$ is the Lorentz factor of $\hat{v}_{s}$ , $r_{I}$ satisfies \eqref{5} and also $R^{\prime \prime}(r)=0$. In this case, the $e_{ \pm}^{\infty}$ of \eqref{12} becomes as \cite{39}
\begin{equation}
\begin{aligned}
e_{ \pm}^{\infty}&=\alpha \hat{\gamma}_{s} \gamma_{\text {out }}\left[\left(1+\beta^{\varphi} \hat{v}_{s}^{(\varphi)}\right) \pm v_{\text {out }}\left(\hat{v}_{s}+\beta^{\varphi} \frac{\hat{v}_{s}^{(\varphi)}}{\hat{v}_{s}}\right) \cos \xi \mp v_{\text {out }} \beta^{\varphi} \frac{\hat{v}_{s}^{(r)}}{\hat{\gamma}_{s} \hat{v}_{s}} \sin \xi\right]\\&-\frac{\alpha}{4 \hat{\gamma}_{s} \gamma_{\text {out }}\left(1 \pm \hat{v}_{s} v_{\text {out }} \cos \xi\right)},
\end{aligned}
\end{equation}
where $v_{\text {out }}$ is the outflow speed and $\gamma_{\text {out }}$ is its Lorentz factor, which can be expressed as 
\begin{equation}
v_{\text {out }}=\sqrt{\frac{\sigma}{\sigma+1}}, \gamma_{\text {out }}=\sqrt{1+\sigma}.
\end{equation}

Following the approach for circular orbits, we plot the allowed energy extraction region in the ${r}-{a}$ plane for the plunging case in Figs. 17-20. The red line represents the  event horizon, blue line represents the  ergosphere, purple line represents the  photon sphere radius, and black line represents the  ISCO. From left to right, $\sigma=100,30,10,3$.

\begin{figure}[!h]
\centering
\includegraphics[width=0.99\linewidth]{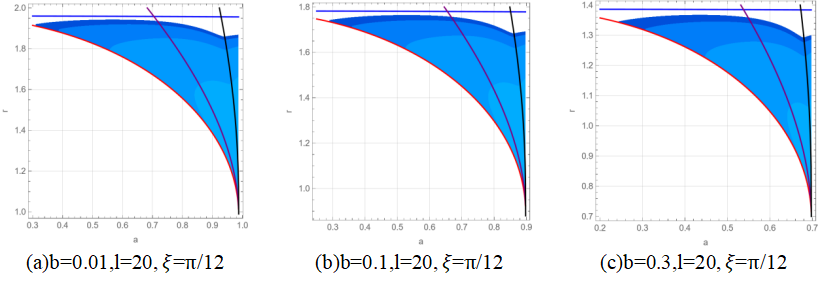}
\caption{ Allowed energy extraction regions for different $b$   in the plunging region.}
\label{fig:17}
\end{figure}

From Fig. 17(a), we see that as $\sigma$ increases, the allowed energy extraction region also increases. Comparing with Fig. 5(a), for $r > r_{I}$ , the allowed region matches the circular orbit case. For $r < r_{I}$, the allowed region in the plunging region is much larger than in circular orbits. Additionally, the minimum spin allowing energy extraction in the plunging region is lower than in circular orbits, and the position of the reconnection point is higher. This is consistent with \cite{26}. For example, taking $\sigma=100$, the minimum spin for energy extraction in circular orbits (Fig. 5(a)) is 0.85, while in the plunging region, it is 0.31. Comparing the three panels of Fig. 17, we see that as $b$ increases, the minimum spin allowing energy extraction also decreases: from 0.31 in Fig. 17(a) to 0.25 in Fig. 17(c). This enables energy extraction from slowly rotating black holes. 

\begin{figure}[!ht]
\centering
\includegraphics[width=0.99\linewidth]{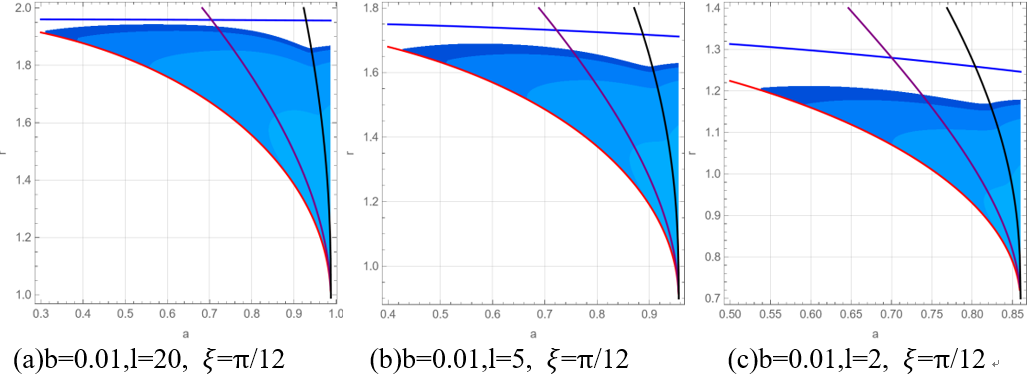}
\caption{Allowed energy extraction regions for different $l$  in the plunging region.}
\label{fig:18}
\end{figure}

From Fig. 18, we see that as $l$ decreases, the minimum spin allowing energy extraction increases. This is inconsistent with the trend in the circular orbit region. Therefore, we conclude that  in the plunging region, $b$ promotes lowering the energy extraction spin threshold, while $l$ suppresses it. However, even with this, the minimum spin in the plunging region is still lower than in circular orbits. For example, taking $\sigma=100,l=2$ (Fig. 18(c)), the minimum plunging region spin is 0.54, while for circular orbits (Fig. 6(c)),  it is 0.795. This highlights the advantage of the plunging region.

\begin{figure}[!ht]
\centering
\includegraphics[width=0.99\linewidth]{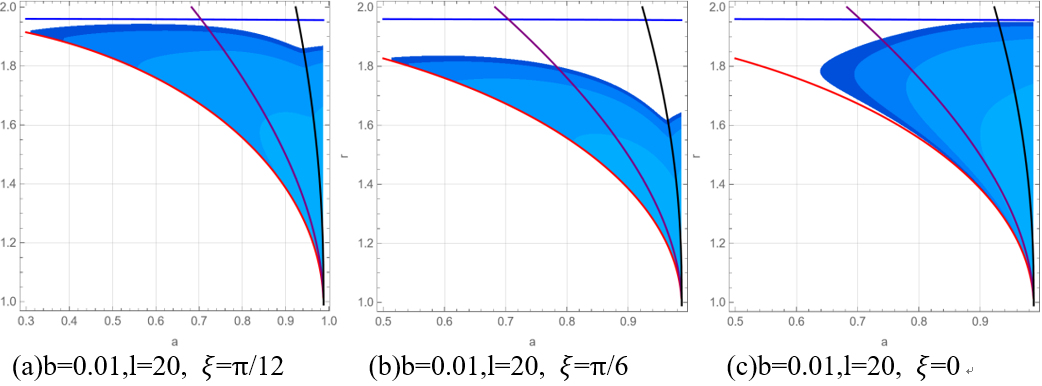}
\caption{Allowed energy extraction regions for different $\xi$ values in the plunging region.}
\label{fig:19}
\end{figure}

From  Fig. 19, we see that the shape of the allowed region changes with the azimuthal angle, which is a characteristic feature of the plunging region azimuthal dependence, which is consistent with  \cite{26}.

\begin{figure}[!ht]
\centering
\includegraphics[width=0.4\linewidth]{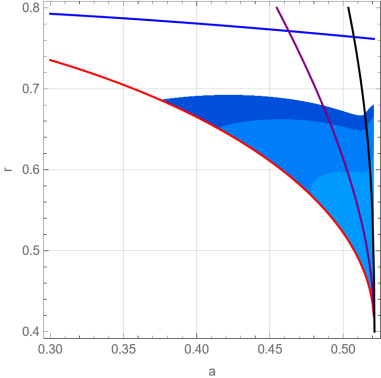}
\caption{Allowed energy extraction region for ${b}=0.36,l=2, \xi=\pi / 12$ in the plunging region.}
\label{fig:20}
\end{figure}

Following Fig. 8, we plot the allowed region for large $b$ and small $l$ in Fig. 20. Here, the minimum spin for energy extraction at $\sigma=100$ is 0.375. If $l$=20, then $b=0.36 > 0.3$ would leads to  a spin less than 0.25 based on previous patterns, but this is not the case here because the presence of $l$ raises the spin threshold.

\subsection{Power and Efficiency of Energy Extraction in the Plunging Region}

Similarly, we plot the energy extraction power in Fig. 21, still using \eqref{15}. To highlight the difference, we plot power for both plunging and circular orbit cases within $r < r_{I}$. We take ${b}=0.01,l=20, \xi=\pi / 12, {a}=0.9, \sigma=100$, here all $r$ values are less than ISCO.

\begin{figure}[!h]
\centering
\includegraphics[width=0.5\linewidth]{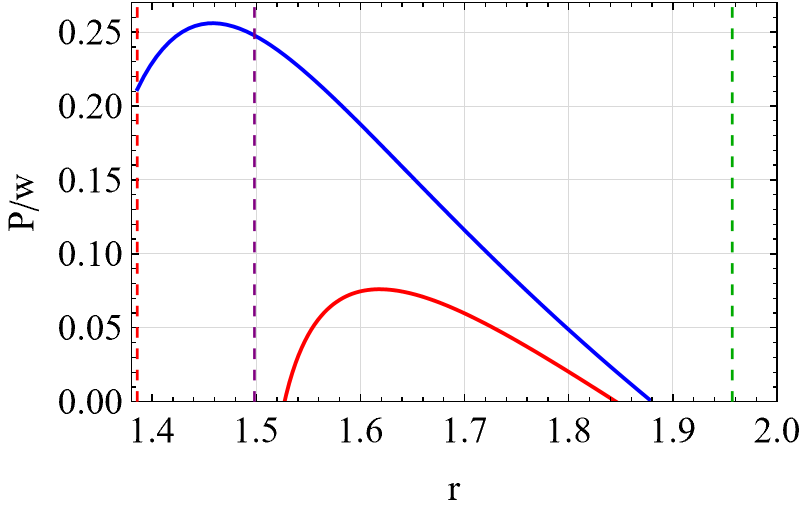}
\caption{Red dashed, purple dashed, green dashed lines represent event horizon, photon sphere radius, ergosphere, respectively. Red solid line represents the  circular orbit power and blue solid line represents the plunging region power.}
\label{fig:21}
\end{figure}

From Fig. 21, it can be seen that the energy extraction power in the plunging region is higher than that in the circular orbit region. Similarly, we compare the energy extraction efficiency for both cases using \eqref{17}, shown in Fig. 22. We use the same parameters in   Fig. 21 and  Fig. 22, that is  ${b}=0.01,l=20, \xi=\pi / 12, {a}=0.9, \sigma=100$.

\begin{figure}[!ht]
\centering
\includegraphics[width=0.5\linewidth]{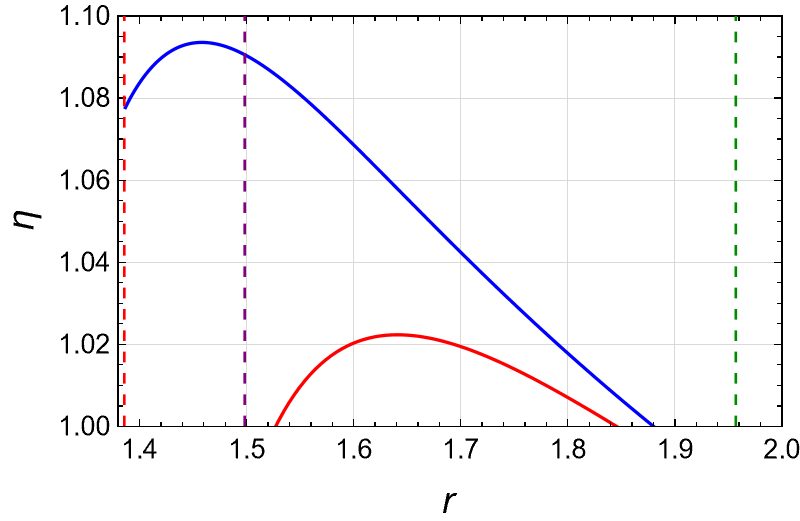}
\caption{Red dashed, purple dashed, green dashed lines represent event horizon, photon sphere radius, ergosphere, respectively. Red solid line represents the  circular orbit 
 efficiency and blue solid line represents the plunging region efficiency.}
\label{fig:22}
\end{figure}

From Fig. 22, the energy extraction efficiency in the plunging region is higher than in the circular orbit region. Comparing with Fig. 21, we see that for the same parameters, the trend of power change is similar to the trend of efficiency change.

Finally, we calculate the power ratio for the plunging region compared to the BZ process, still using \eqref{21}. Using the same parameters as the circular orbit high spin case (${b}=0.01,  1=20,  \sigma=3, {a}=0.98,  {r}=1.2$,  $\xi=\pi / 12$), and noting that $r_{I}=1.379$  > $r$ satisfies the plunging region condition, we get a power ratio of 15.6634 > 10.3056. This again shows that plunging region power exceeds both circular orbit power and BZ power. Now considering a low-spin case based on Fig. 17(c) (${b}=0.3,1=20, \xi=\pi / 12$), spin $a$ can be 0.3 or lower. However, note that here the photon sphere radius is above the ergosphere, so \eqref{16} is no longer valid and should be replaced by
\begin{equation}
A_{i n} \sim\left(r_{E}^{2}-r_{+}^{2}\right). 
\end{equation}
Taking $\sigma=100, {a}=0.3, {r}=1.34, \xi=\pi / 12$, we get a power ratio of 0.0393679, which is very low. This is understandable because in the plunging region, to achieve low spins, $\sigma$ must be large. According to \eqref{21}, $\sigma$ appears in the denominator, reducing the power ratio. Nevertheless, the ability to extract energy via magnetic reconnection at spins as low as 0.25 is very interesting, for previous studies rarely achieved such low spins.

\section{Conclusion}
Magnetic reconnection provides a new mechanism for extracting energy from rotating black holes. How to extract energy efficiently and conveniently is the  goal pursued by physicists. Centering on this goal, we conducted an investigation in  Kerr-Sen-AdS$_{4}$ black hole.  We discussed not only in   the circular orbit region  but also in the plunging region. First, we introduced the event horizon, boundary of ergosphere, and photon sphere radius of this black hole. We found that due to the presence of the dilatonic scalar charge $b$ and AdS radius $l$, the maximum allowed spin is reduced compared to the Kerr black hole. Then, we described the energy extraction mechanism via magnetic reconnection in circular orbits. We analyzed the feasibility of energy extraction for this black hole, plotted the parameter space, and analyzed the influence of parameters $b$, $l$, $\xi$ and $\sigma$ on energy extraction.  We also analyzed the energy extraction power and efficiency, especially under the influence of $b$ and $l$, and compared them with the BZ mechanism, finding that magnetic reconnection power can exceed BZ power. Finally, we studied energy extraction in the plunging region, describing the magnetic reconnection mechanism there. We continued analyzing the influence of parameters $b$, $l$, $\xi$ and $\sigma$, comparing the results with the circular orbit case. We found that the power and efficiency of energy extraction are higher in the plunging region.

For circular orbits, we found that energy extraction is possible even at a spin of 0.5. For the plunging region, we found that energy extraction is possible even at a spin of 0.25. This is caused by the influence of the two parameters $b$ and $l$ in this black hole. We also found that in circular orbits, both $b$ and $l$ promote lowering the spin threshold for energy extraction. However, in the plunging region, the situation is different, that is,  $b$ promotes lowering the spin threshold, while $l$ suppresses it.

This conclusion indicates that the plunging region has a stronger energy extraction capability. To extract energy, it is best to operate in the plunging region. For circular orbits, increasing $b$ or decreasing $l$ enables energy extraction at lower spins. For the plunging region, increasing $b$ or increasing $l$   enables energy extraction at lower spins.

\noindent {\bf Acknowledgments}

\noindent
This work is supported by the National Natural Science Foundation of China (Grants No. 12375043).

\end{document}